# Design Kinetic Parameters for Improved Resilience of Materials under Irradiation

M. Nahavandian[1], E. Aydogan[2], J. Byggmästar[3], M.A. Tunes[4], E. Martinez[1,5*], O. El-Atwani[2*]

[1]School of Mechanical and Automotive Engineering, Clemson University, Clemson, SC, 29634, USA

[2]Reactor Materials Group, Nuclear Sciences Division, Pacific Northwest National Laboratory, Richland WA, 99352, USA

[3]Department of Physics, University of Helsinki, FI-00014 Helsinki, Finland

[4]Department of Metallurgy, Chair of Nonferrous Metallurgy, Montanuniversität Leoben, 8700 Leoben, Austria

[5]Department of Materials Science and Engineering, Clemson University, Clemson, SC, 29634, USA

**Abstract:** High entropy alloys (HEAs) have captured much attention in recent years due to their conceivably improved radiation resistance compared to pure metals and traditional alloys. However, among HEAs, there are millions of design possibilities considering all potential compositions. In this study, we develop criteria to design HEAs with improved radiation resilience taking into consideration defect properties to promote interstitial-vacancy recombination. First, we conduct rate theory calculations on defects followed by molecular dynamics (MD) simulations on pure W and W-based multicomponent concentrated alloys. It is found that when the diffusion coefficients for single vacancies and interstitials become similar and the effective migration energies of defects is minimum (maximum diffusivities), defect recombination becomes optimal, and the concentration of defects is significantly reduced. This is supported by MD simulations indicating improved radiation resistance of V- and Cr-based alloys, which satisfy the above-stated criteria. Furthermore, experimental observations also reinforce the proposed approach. This study sheds light on the design criteria for improved radiation resistance and helps material selection without the need of extensive experimental work.

**Keywords:** High entropy alloys, radiation resistance, rate theory, molecular dynamics

* Corresponding author: oelatwan25@gmail.com, enrique@clemson.edu

**Introduction**

Successful deployment of next-generation nuclear reactors hinges on advanced, high-performance structural materials that can safely withstand extremely hostile environments for prolonged periods of time.[1] These environments might include extreme combinations of high temperatures, large time-varying stresses, high dose neutron fluxes and helium generation (He), introduced as a transmutation or spallation product. Neutron irradiation creates interstitial and vacancy point defects in excess concentrations. These defects may evolve into extended defects such as voids, dislocation loops and stacking fault tetrahedra, causing deterioration of the dimensional and structural stability of the materials.[1] The radiation response of materials is hence closely related to the annihilation, coalescence and recombination of these defects. While the first one requires defect sinks, the latter two can be mainly controlled by defect properties.[2] Thus, for improved radiation resistance, nano-layered and nanograined pure materials and alloys having high density of sinks have been explored extensively.[3-5] However, irradiation-induced grain growth and instabilities in the microstructure at high temperatures pose problems for conventional materials.[6] Another route of improving the radiation resistance of materials is alloying to alter defect evolution and enhance defect recombination.[8] For instance, Cr addition in Fe has been reported to improve the radiation resistance.[9] Similarly, increasing the Cu content in Ni–Cu alloys may effectively suppress irradiation-induced void swelling at the peak swelling temperature.[10]

      A new class of materials, compositionally complex alloys (CCAs) which include medium or high entropy alloys (MEA or HEA), composed of multiple elements with a concentration between 5 and 35 at% have started to attract much attention due to their potentially improved radiation resistance compared to the conventional alloys.[2,11] The local lattice distortion, complexities from random arrangement of alloying elements, and the intrinsically complex local chemical environment can alter formation energies, migration barriers and diffusion pathways of irradiation-induced defects, promoting interstitial-vacancy recombination, and resulting in improved radiation resistance.[2,11] In their review paper, Cheng et al.[12] classified the radiation damage as radiation-induced hardening, void swelling, radiation-induced segregation/precipitation, and helium behavior. They concluded that CCAs are generally more radiation resistant compared to conventional alloys, with BCC CCAs being more resistant compared to FCC CCAs, presumably due to their larger stacking fault energies, greater lattice

distortion and lower defect generation rates. For instance, Kareer et al.[13] reported a much less radiation-induced hardening in TiVTaZr, TiVTaCr and TiVTaNb compared to pure V after ion irradiation at 500 °C up to a peak dose of 3.6 dpa. Similarly, El-Atwani et al.[14,15] investigated the radiation response of WTaVCr alloys using heavy ions and He implantation and compared the results with W-based alloys. They reported that there is no dislocation loop formation or considerable hardening after heavy ion irradiation, and significantly reduced change in volume due to He bubbles after He implantation.[15] However, to prevent Cr-V precipitation after heavy ion irradiation, Hf was added into the original WTaCrV composition to obtain single crystalline CCA with high thermal stability and radiation resistance even under dual beam irradiations, supported by cluster expansion and density-functional theory calculations.[16] Qiu et al.[17] studied the primary radiation damage and long-term defect evolution in pure V, V-5Ti-5Ta conventional alloy, VTiTa MEA and VTiTaNb HEA using molecular dynamics simulations. MEAs and HEAs exhibited lower defect clustering fraction as well as smaller cluster and dislocation loop size compared to pure V and V-5Ti-5Ta alloy because of lower dislocation loop binding energy and defect mobility.

However, there is an infinite number of possible compositions to design new CCAs. Elemental selection can be done based on the application, such as radiation resistance, high temperature stability, oxidation resistance, low density materials, high melting temperatures, thermal conductivities, etc.[14,16,18-20] Several studies employed empirical rules for solid solution formation, CALPHAD calculations, cluster expansion and machine-learning.[14,16,18-21] Here, we investigate the criteria that may lead to optimal defect recombination, and thus potential improvement in resilience to irradiation, based on rate theory calculations and molecular dynamics simulations. These findings are supported by experimental data proving that these criteria should broadly lead to enhanced radiation resistance.

**Results and Discussion**

The radiation resistance of the HEAs is mostly related to four core effects, namely high mixing entropy, cocktail effect, sluggish diffusion, and severe lattice distortion.[22] It is attributed to different factors: (i) reduced thermal conductivity due to disorder, which extends the thermal phase of the cascade promoting recombination,[2,11,23] and (ii) distinct defect properties due to chemical disorder, lattice distortion and sluggish diffusion, which result in peculiar diffusion through rough energy landscapes, thus altering defect migration properties potentially enhancing recombination.[2,24,25] Several studies demonstrated different radiation damage evolutions in HEAs compared to pure materials and other alloys.[2,17,23,26,27] This indeed infers the change in defect properties, which is embedded in rate theory equations. To demonstrate this, we have conducted rate theory calculations for single defects, varying migration energies for vacancies and self-interstitials independently, and temperature. Figure 1 shows that when the diffusion coefficient of vacancies and interstitials ($D_v$ and $D_i$, respectively) are the same and their total migration energy ($\Delta E_v + \Delta E_i$) is small, the total concentration of vacancies and interstitials in the matrix ($C_v+C_i$) is also low, indicating high radiation resistance. The calculations shown in Fig. 1 were performed at one selected temperature (500 K) to demonstrate the effect of having different migration energies between interstitials and vacancies. Defect properties in different material systems were adopted from the literature and superimposed on the plot in Fig. 1.[27] According to these findings, W has the highest density of defects due to the largest difference in the diffusion coefficients of vacancies and interstitials, with a very large migration energy for vacancies ($\Delta E_v \sim$ 1.54 eV) and low for interstitials ($\Delta E_i \sim$ 0.2 eV). Similarly, WMo has both large total migration energies and large differences in diffusivities of vacancies and interstitials. WTa has smaller ratio between diffusivities of defects, leading to smaller concentration of defects. WTaVMo, WTaV and WV have $D_i/D_v$ ratios of ~1 and small total migration energy, resulting in low defect concentration (WV being the lowest). It should be noted that the more complex system WTaVMoNb has high concentration of defects due to higher diffusivity of vacancies than that of interstitials and high migration energies of both defect types.

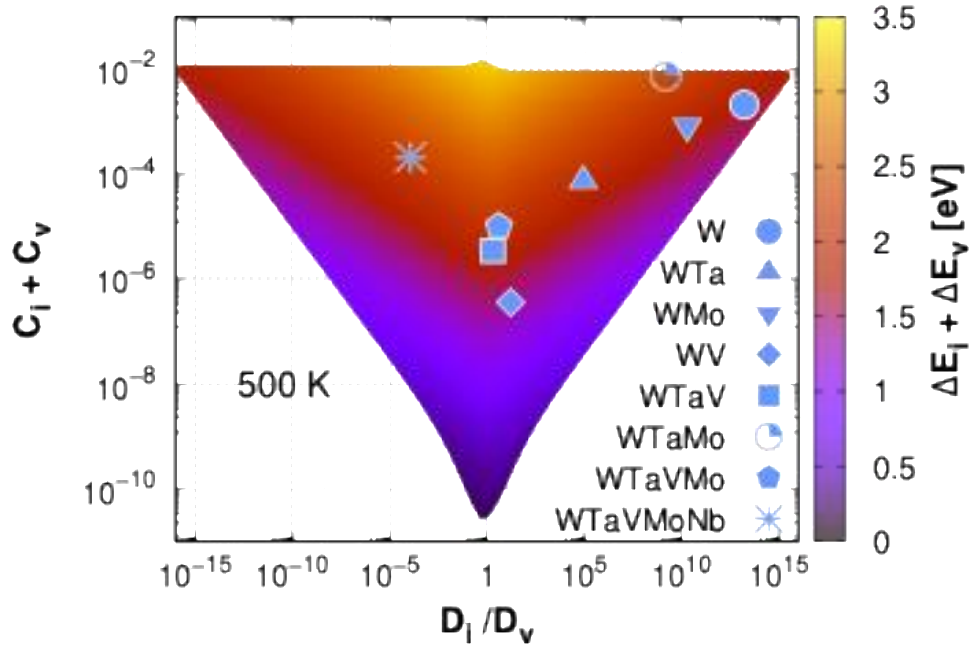

**Figure 1**. Defect concentrations as a function of diffusivity ratio and migration energies at 500K. Blue symbols denote defect properties from literature [27]. The figure shows which defect properties (selection of elements) should give the highest irradiation resistance.

Besides low concentration of single interstitial and vacancy defects, recombination and coalescence rates of individual defects will also determine the final radiation resistance of the material. Figure 2 shows the rate theory results, as a function of temperature, including the propensity for clustering. When the ratio of total diffusion coefficient of vacancies and interstitials ($D_i + D_v$), as related to the recombination rate, to the diffusion coefficient for defect coalescence for both interstitials ($D_i + D_i$) and vacancies ($D_v + D_v$), as related to the clustering rates, becomes 1, the material achieves the optimal radiation resistance. Defect properties of W, WTa, WMo, WV, WTaV, WTaMo, WTaVMo and WTaVMoNb in Ref [27] are superimposed on Fig. 2. Based on the results, especially at low temperatures, W has the highest defect density ($C_i + C_v$), followed by WMo and WTaMo due to the large difference between the diffusivities of vacancies and interstitials, which results in large diffusivity ratios between recombination and vacancy clustering. Similarly, since the diffusivity of vacancies is higher than that of interstitials in WTaVMoNb, the defect concentration is high due to the large diffusivity ratio of recombination to interstitial clustering. As the temperature increases, the rate for vacancy cluster formation becomes comparable to that of recombination and the ratios become smaller (closer to 1). However, at all temperatures, WTaV and WV have the diffusion coefficient ratios the closest to 1, and therefore, the lowest defect concentration. It should be noted that the model does not consider damage cascade, or any time-dependent evolution of sinks.

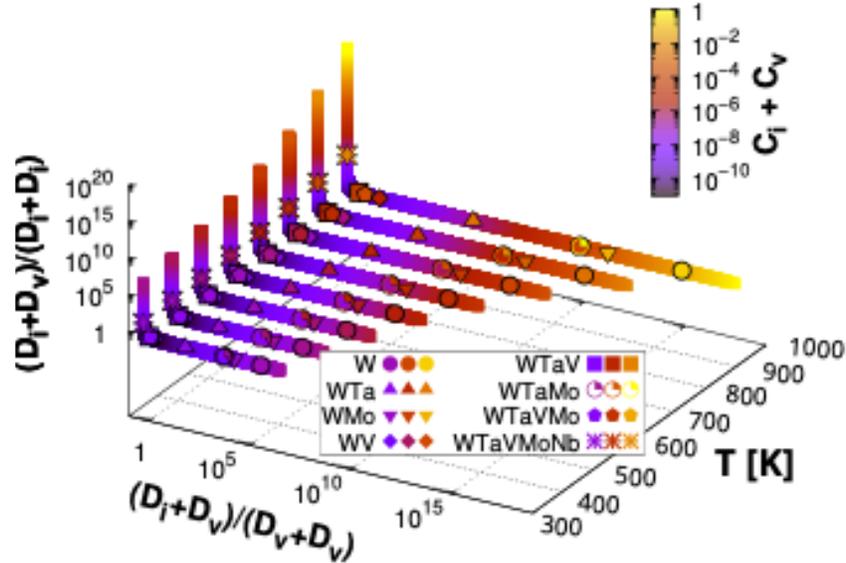

**Figure 2.** 4D figure of recombination to defect coalescence ratios and defect concentrations as a function of temperature. The values of W and some other alloys are adopted from Ref [27].

To support the rate theory findings, MD annealing simulations were conducted at 2000 K after the introduction of 0.1% and 2% Frenkel pairs in pure W, W binaries, ternaries, quaternaries and quinary systems, as seen in Fig. 3. After introduction of 0.1% Frenkel pairs, W shows the highest number of remnant defects after 10 ns, while defects are mostly annihilated in systems containing Cr and V. The difference in the final defect concentration after annealing is much more dramatic for the case of 2% Frenkel pairs. Figure 3 shows that there is a considerable number of defects remaining in W while alloys containing Ta and/or Mo have a lower number of defects. However, alloys with V and/or Cr, namely WV, WCr, WTaV, WTaCr, and WTaCrV have by far the lowest defect concentration after annealing at 2000 K. The snapshots of the final annealed systems in Figure 3 also show that in W, large interstitial clusters are formed while in WTaV (and other V- and Cr-containing alloys), recombination is so dominant that no clusters survive. These findings support the rate theory results in Figure 2. Zhao[28] conducted first principles calculations on the equiatomic WTaCrV alloy and reported enhanced radiation resistance as a result of pronounced lattice distortions due to a large size difference in the elemental constituents together with the existence of V and Cr dumbbells. Interstitial dumbbells along [110], rather than typical [111] dumbbells in BCC metals, have been determined to contribute to slow interstitial diffusion in this system. Such slower diffusivities lead to a larger overlap in the migration energies of vacancies and interstitials, helping recombine Frenkel pairs.

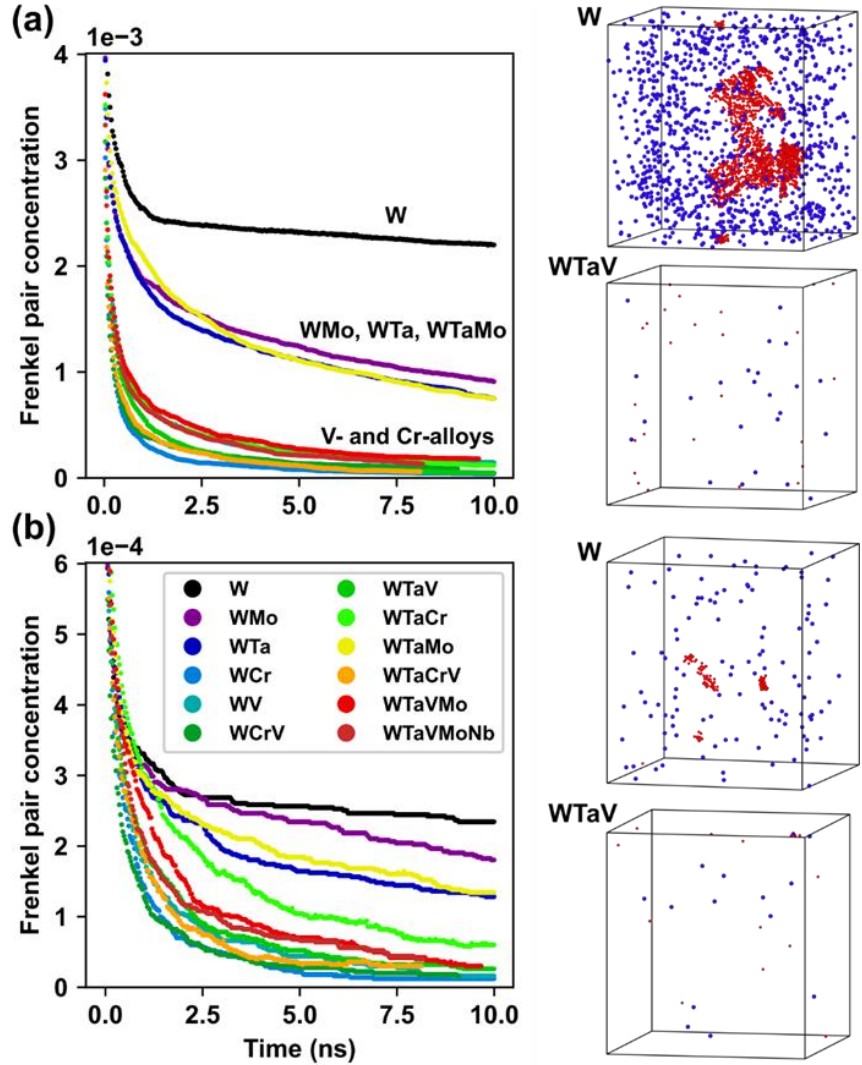

**Figure 3.** Concentration of Frenkel pairs in MD annealing simulations on W, WV, WMo, WTa, WV, WCr, WCrV, WTaV, WTaCr, WTaMo, WTaCrV, WTaVMo and WTaVMoNb systems after inserting (a) 2% and (b) 0.1% of randomly placed Frankel pairs and annealing at 2000 K. The snapshots show the final defects after 10 ns annealing, with blue atoms representing vacancies and red atoms self-interstitial atoms.

The findings from modelling have been verified by experimental data. The micrographs in Fig. 4 show the microstructures of pure W in coarse grained (CG) and ultra fine grained (UFG) conditions, CG W-Ta, UFG WTaV, WTaCrV and WTaCrVHf alloys after heavy ion irradiation and He implantation at elevated temperatures.[14,16,29-32] Defect concentrations decrease with the decrease in grain size in the case of pure W, yet there is considerable amount of dislocation loops and cavities in UFG W, as seen in Figure 4a&g. However, micrographs compiled from the authors' previous studies and the literature clearly show the effect of alloying compared to the density of sinks (grain boundaries). While the CG W shows high amount of dislocation loops as well as high density of cavities after being irradiated at 1050 K to 0.25 dpa with 3 MeV $Cu^+$ (Figure 4b&h), W-5%at Ta alloy has mostly black dots and a lower amount of cavities after irradiation to 1 to 1.2 dpa at 773K using 2 MeV $W^+$ ions (Figure 4c&i). The irradiation damage nevertheless decreases considerably with the increase in compositional complexity. For instance, UFG WTaV, WTaCrV and WTaCrVHf alloys show neither loop damage nor He bubbles or cavities, except minor amount of bubbles forming in the WTaCrVHf alloy, irradiated at 1073 K up to 20, 8 and 8.5 dpa, respectively.

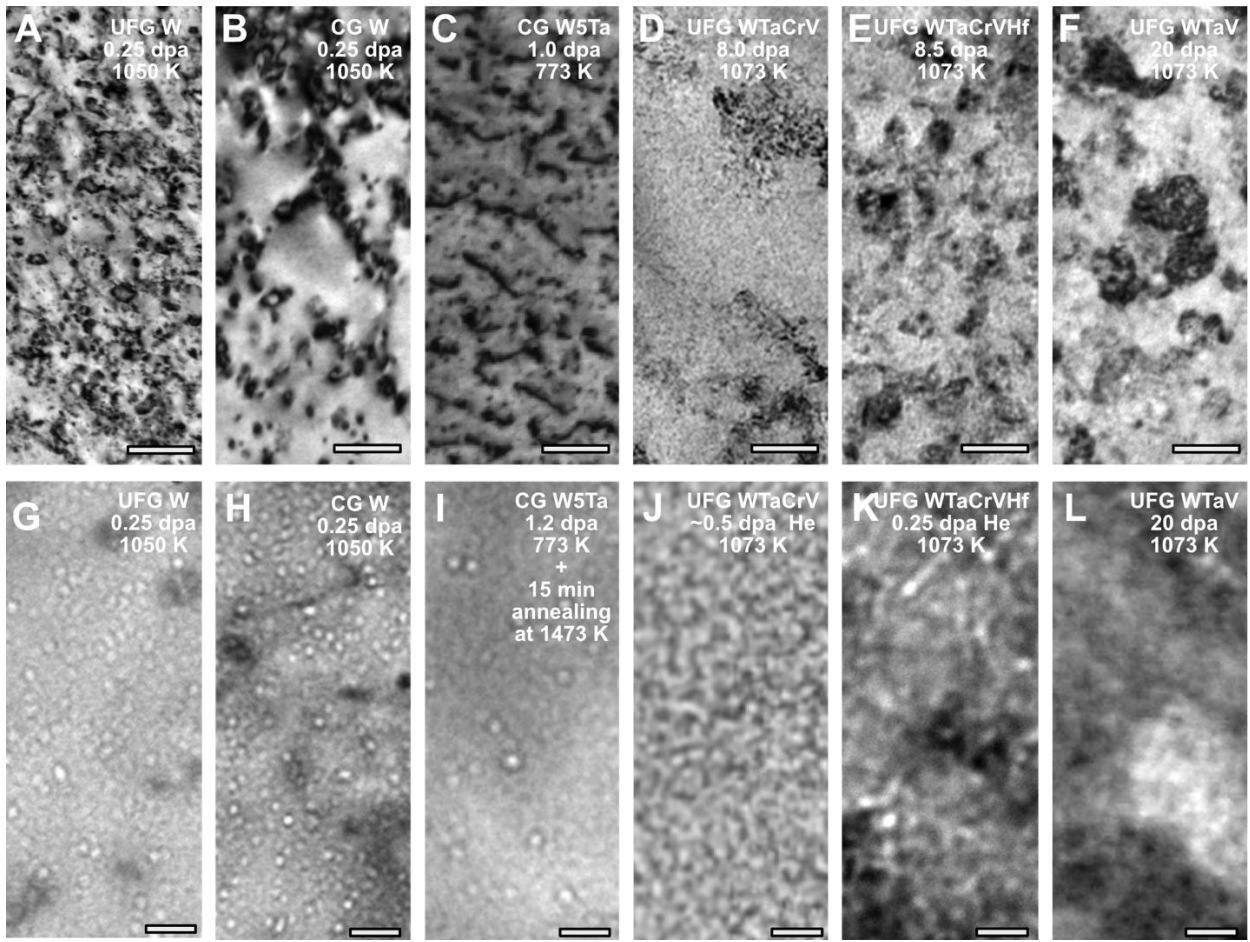

**Figure 4.** Evolution of W-based candidate fusion materials | (A,F) CG W, (B,G) CG WTa, (C,H) WTaCrV, (D,I) WTaCrVHf, and (E,J) WTaV. Scale bar in (A-E) is 50 nm. Scale bar in (F-J) is 10 nm. All micrographs are BFTEM, except J which is HAADF. [14,16,29-32]

He damage evolution, in the microstructure of He implanted materials, was also observed to be dependent on defect properties. The damage in this case is closely related to temperature. When the temperature is between $0.2T_m$ and $0.5T_m$ (with $T_m$ the melting temperature), the ratio of diffusion coefficients of He to vacancy is reported to be in the order of $10^2$ to $10^6$ in the BCC FeCrVTi system.[33] However, due to the sluggish diffusion of interstitials and thus similar diffusion coefficients of vacancies and interstitials, recombination of Frenkel pairs makes the transport of He atoms difficult and results in reduced swelling.[35] For example, El-Atwani et al. investigated the He bubble evolution in WTaCrV system and reported a 2-3 nm size He bubbles uniformly distributed in the microstructure. First principles calculations indicated that the smaller bubble damage evolution in the alloy compared to pure W is due to lower diffusivity of He in the matrix and implied enhanced interstitial and vacancy recombination.[15] It is, therefore, apparent that defect properties in HEAs constitute a dominant factor in their radiation resistance, and the design of HEAs can benefit from the criteria developed in this work.

**Conclusions**

The alloy design scope for high entropy alloys (HEAs) is immense due to the possible combination of many different elements. Even with high-throughput simulation, production and characterization methods, it is impossible to produce and test or model each alloy composition. Thus, setting certain design criteria is helpful for downselecting optimal compositions. This study shows that both rate theory calculations and MD simulations can be used to help generate improved HEAs considering certain criteria for enhanced radiation resistance. These criteria require similar diffusion coefficients for vacancies and interstitials, and minimum total migration energies for the material to achieve optimal radiation resistance to defect accumulation.

**Acknowledgements**

Funding from the U.S. Department of Energy, Advanced Research Projects Agency-Energy under contract DE-AR0001541 is gratefully acknowledged.

**Data Availability**

All data and codes used in the simulations performed in this paper are available upon request.

**Methods**

*Rate theory*

Standard rate theory models[37] describe the evolution of the average vacancy (v) and self-interstitial (i) concentrations under irradiation, $C_v$ and $C_i$, as a set of coupled chemical differential equations:

$$\frac{dC_v}{dt} = G' - K_{iv}C_vC_i - \sum_s k_{sv}^2 D_v C_v \qquad (1)$$

$$\frac{dC_i}{dt} = G' - K_{iv}C_vC_i - \sum_S k_{si}^2 D_i C_i \qquad (2)$$

The first term in the RHS of Eqs. (1) and (2), $G'$, is an effective production rate. The second term denotes the recombination rate, derived from Waite's theory of the kinetics of diffusion-limited reactions [38]:

$$K_{IV} = 4\pi R_{rec} \frac{D_i + D_v}{\Omega}$$

where $D_i$ and $D_v$ are the diffusion coefficients of point defects, $R_{rec}$ is the recombination distance taken as $3a_0$ with $a_0 = 0.3165$ nm the lattice parameter, and $\Omega$ the atomic volume.

The last terms in Eqs.(1) and (2), $\sum_s k_{sv}^2 D_v C_v$ and $\sum_s k_{si}^2 D_i C_i$, are the rates of elimination at point defect sinks. Each kind of sink $s$ is characterized by its sink strength $k_{sv}^2$ and $k_{si}^2$ [39]. If sink biases are neglected and $k_{sv}^2 = k_{si}^2$, Eqs.(1) and (2) become

$$\frac{dC_v}{dt} = G' - K_{iv}C_vC_i - K_v C_v \qquad (1)$$

$$\frac{dC_i}{dt} = G' - K_{iv}C_vC_i - K_i C_i \qquad (2)$$

with $K_d = \sum_s k_{sd}^2 D_d$.

At steady-state the average concentration of defects in the domain is given by

$$C_v^{st} = -\frac{K_i}{2K_{iv}} + \left[\left(\frac{K_i}{2K_{iv}}\right)^2 + \frac{G'K_i}{K_{iv}K_v}\right]^{1/2}$$

$$C_i^{st} = -\frac{K_v}{2K_{iv}} + \left[\left(\frac{K_v}{2K_{iv}}\right)^2 + \frac{G'K_v}{K_{iv}K_i}\right]^{1/2}$$

These steady-state concentrations depend strongly on defect diffusivities, which have been estimated following harmonic transition state theory as [40]

$$D_v = \nu a^2 \exp\left(-\frac{\Delta E_d}{k_B T}\right), \qquad D_i = \frac{1}{2}\nu a^2 \exp\left(-\frac{\Delta E_d}{k_B T}\right)$$

Where $\nu$ is an attempt frequency, $a$ the lattice parameter, $\Delta E_d$ the migration energy of defect $d$, $k_B$ the Boltzmann constant and $T$ the temperature. $D_i$ assumes a Johnson mechanism in BCC with the self-interstitial in a <110> configuration exchanging with one of four nearest neighbors. We have taken $\nu = 10^{13}$ s$^{-1}$ for both defects and we have varied the energy barriers from 0.2 to 1.65 eV in increments of 0.01 eV and the temperature from 300 K to 1000 K in 100 K increments.

*Molecular Dynamics*

To investigate the efficiency of clustering versus recombination of defects we performed molecular dynamics simulations using state-of-the-art machine-learned interatomic potentials. With pure W as a reference, we chose a set of refractory alloys from binaries to quinaries and simulated the evolution of randomly placed vacancies and self-interstitial atoms at 2000 K and zero pressure. All alloys are simulated at equiatomic composition and randomly ordered in systems of half million atoms. The simulations were done with the LAMMPS code[41] using tabGAP machine-learned potentials.[42] For each alloy we prepared two initial systems, one with a supersaturated (2%) Frenkel pair concentration and one with a more realistic concentration of 0.1%. During the annealing simulations, the interstitials and vacancies were identified with the Wigner-Seitz method[43] and visualized with OVITO [44].